\documentclass[aps,prl,10pt,amsmath,floats,floatfix,twocolumn,superscriptaddress,altaffilletter,nofootinbib,shownopacs,longbibliography]{revtex4-1}
\newcommand\prlsec[1]{\vspace{2mm}\noindent \textbf{\emph{#1}}\,---}
\usepackage[T1]{fontenc}
\usepackage[utf8]{inputenc}
\usepackage{lmodern}
\usepackage{verbatim}
\usepackage{physics}
\usepackage[dvipsnames, usenames]{xcolor}
\definecolor{linkcolor}{rgb}{0.6,0.0,0.0}
\definecolor{applegreen}{rgb}{0.55, 0.71, 0.0}
\usepackage{appendix}
\usepackage{booktabs}
\usepackage{enumitem}
\usepackage[hypertexnames=false, unicode, colorlinks=true, linkcolor=linkcolor,
citecolor=linkcolor, filecolor=linkcolor,urlcolor=linkcolor,
pdfusetitle]{hyperref}

\usepackage[all]{hypcap}
\usepackage{graphicx}
\usepackage{color}
\usepackage{xspace}
\usepackage{amssymb}
\usepackage[normalem]{ulem} 
\usepackage{bm} 
\usepackage{orcidlink}
\usepackage{aasmacros}
\usepackage{microtype}
\usepackage[english]{babel}
\usepackage{blindtext}

\graphicspath{%
  {figs/}%
  {figs/EO}
  {figs/PP}
  {figs/CO}
  {figs/BW}
}

\def\beq{\begin{equation}\begin{aligned}}
\def\eeq{\end{aligned}\end{equation}}

\begin{document}

\newcommand{\IUCAA}{\affiliation{Inter University Center for Astronomy and Astrophysics, Ganeshkhind, Pune 411007, India}}
\newcommand{\IISERPune}{\affiliation{Indian Institute of Science Education and Research, Homi Bhabha Road, Pashan, Pune 411008, India}}
\newcommand{\TIFR}{\affiliation{Tata Institute of Fundamental Research, Homi Bhabha Road, Colaba, Mumbai 400005, India}}
\newcommand{\CITA}{\affiliation{Canadian Institute for Theoretical Astrophysics, University of Toronto, 60 St. George Street, Toronto, ON M5S 3H8, Canada}}

\title{Profiling Dark Matter Spikes with Gravitational Waves from Accelerated Binaries}

\author{Avinash Tiwari\,$^*$\,\orcidlink{0000-0001-7197-8899}}
\email{avinash.tiwari@iucaa.in}
\IUCAA{}
\author{~Prolay Chanda\,$^*$\,\orcidlink{0000-0002-3940-6062}}
\email{prolay.chanda@tifr.res.in}
\TIFR
\author{~Shasvath J. Kapadia\,\orcidlink{0000-0001-5318-1253}}
\email{shasvath.kapadia@iucaa.in}
\IUCAA{}
\author{Susmita Adhikari\,\orcidlink{0000-0002-0298-4432}}
\email{susmita@iiserpune.ac.in}
\IISERPune{}
\author{Aditya Vijaykumar\,\orcidlink{0000-0002-4103-0666}\,}
\email{aditya@utoronto.ca}
\CITA{}
\author{Basudeb Dasgupta\,\orcidlink{0000-0001-6714-0014}\,}
\email{bdasgupta@theory.tifr.res.in}
\TIFR{}

\date{August 5, 2025;~{Preprint No.\,TIFR/TH/25-17}}

\begin{abstract}
Dark matter halos can develop a density spike, e.g., around a galactic supermassive black hole, with the profile $\rho \propto r^{-\gamma_{\rm sp}}$ determined both by the galaxy's formation history and the microphysics of dark matter. We show that future LISA/DECIGO observations, of intermediate/stellar-mass binary mergers inside the spike around the supermassive black hole, can measure~$\gamma_{\rm sp}$ at a few-percent--level precision. The spike induces a distinctive time-dependent acceleration along the non-circular orbit taken by the binary's center of mass, which is observable as a secular modulation of the gravitational wave signal. This method --- insensitive to confounding astrophysical effects (dynamical friction, tidal effects, etc.) and not reliant on unknown dark matter particle physics --- provides a clean diagnostic of density spikes and a new probe of dark matter.

\end{abstract}

\maketitle

\renewcommand{\thefootnote}{\fnsymbol{footnote}}
\footnotetext[1]{The first two authors contributed equally to this work.}

\prlsec{Introduction.} Clustering on cosmological scales is a powerful probe of dark matter (DM). The relevant observables, e.g., power spectrum, halo counts \&  shapes, etc., are sensitive to  DM properties. Remarkably, on galactic scales as well, robust and distinctive signatures can point to broad classes of DM through its transport coefficients. One such feature is the formation of a steep DM density spike, with $\rho(r)\propto r^{-\gamma_{\rm sp}}$, that encodes the astrophysical history of the galaxy and yet to be discovered particle properties of DM. 

Several astrophysical scenarios, e.g., adiabatic growth around a black hole (BH)~\cite{Gondolo:1999ef}, secondary infall onto a primordial black hole (PBH)~\cite{bertschinger1985}, and survival of ``prompt cusps'' of DM halos~\cite{Delos:2019mxl}, predict density spikes with characteristic features. From a particle perspective as well, cold and collisionless DM candidates, e.g., WIMPs (and its variants) or axions (or axion-like particles), naturally form a steep spike; whereas those with large de Broglie wavelengths, e.g., fuzzy DM~\cite{Hu:2000ke}, or with significant self-interactions, e.g., self-interacting DM (SIDM)~\cite{Spergel:1999mh}, or with large thermal velocities, e.g., warm DM~\cite{Colin:2000dn}, can suppress spike formation. Gravothermal catastrophe in certain self-interaction models also create spikes \cite{Balberg:2002ue}. Detecting a spike, especially with a precise measurement of its parameters, may therefore reveal important clues.

Detecting DM density spikes remains a formidable challenge. Particle-physics–based approaches, originating with the proposal of Gondolo and Silk~\cite{Gondolo:1999ef}, rely on specific properties of the assumed DM candidate, such as WIMPs~\cite{Ullio:2001fb}, axions~\cite{Edwards:2019tzf,Hannuksela:2019vip}, or SIDM~\cite{Banik:2025fnc}, and are model-dependent. On the gravitational wave (GW) front, the seminal work of Eda et al.~\cite{Eda:2013gg} showed that stellar mass objects merging with an intermediate mass black hole (IMBH) can be used to profile the surrounding DM spike, via phase shifts induced by its gravitational potential. Subsequent analysis revealed that, for such intermediate mass-ratio inspirals (IMRIs), the dominant GW modulation typically arises from dynamical friction (DF) exerted by the halo~\cite{Eda:2014uha}, with accretion and resonances introducing additional dephasing~\cite{Macedo:2013jja}. Subsequent work on this IMRI scenario has therefore focused on modeling the dissipative effects~\mbox{\cite{Barausse:2014pra,Barausse:2014tra,Yue:2017iwc,Yue:2019ozq,Kavanagh:2020cfn,Coogan:2021uqv,Speeney:2022ryg,Tahelyani:2024cvk,Feng:2025fkc}}. More recently, stellar-orbit astrometry~\cite{Lacroix:2018zmg} and reverberation mapping~\cite{Sharma:2025ynw} have been explored, though their sensitivity to the spike index remains limited and subject to degeneracies with other astrophysical parameters. A method capable of inferring the spike profile while minimizing reliance on particle physics or complex astrophysical modeling is therefore still highly desirable.

In this \emph{Letter}, we show that future GW observation of a binary merger within the spike would allow a precise measurement of the spike index $\gamma_{\rm sp}$. The main idea is that the motion of the center of mass (CoM) of a compact binary coalescence (CBC), i.e., its outer orbit, is determined dominantly by the gravitational potential modified by the DM density spike, with negligible effects of dissipative processes. The kinematic parameters (acceleration, jerk, snap, etc.) of the outer orbit imprint themselves as long-term modulations on GWs emitted by the coalescence of the binary; extracting these parameters from the GW signal allows for a precise measurement of $\gamma_{\rm sp}$ --- free of complex astrophysics and confounding degeneracies. We now present our method and results in detail.

\begin{figure}[t]
  \centering
\includegraphics[width=0.9\linewidth]{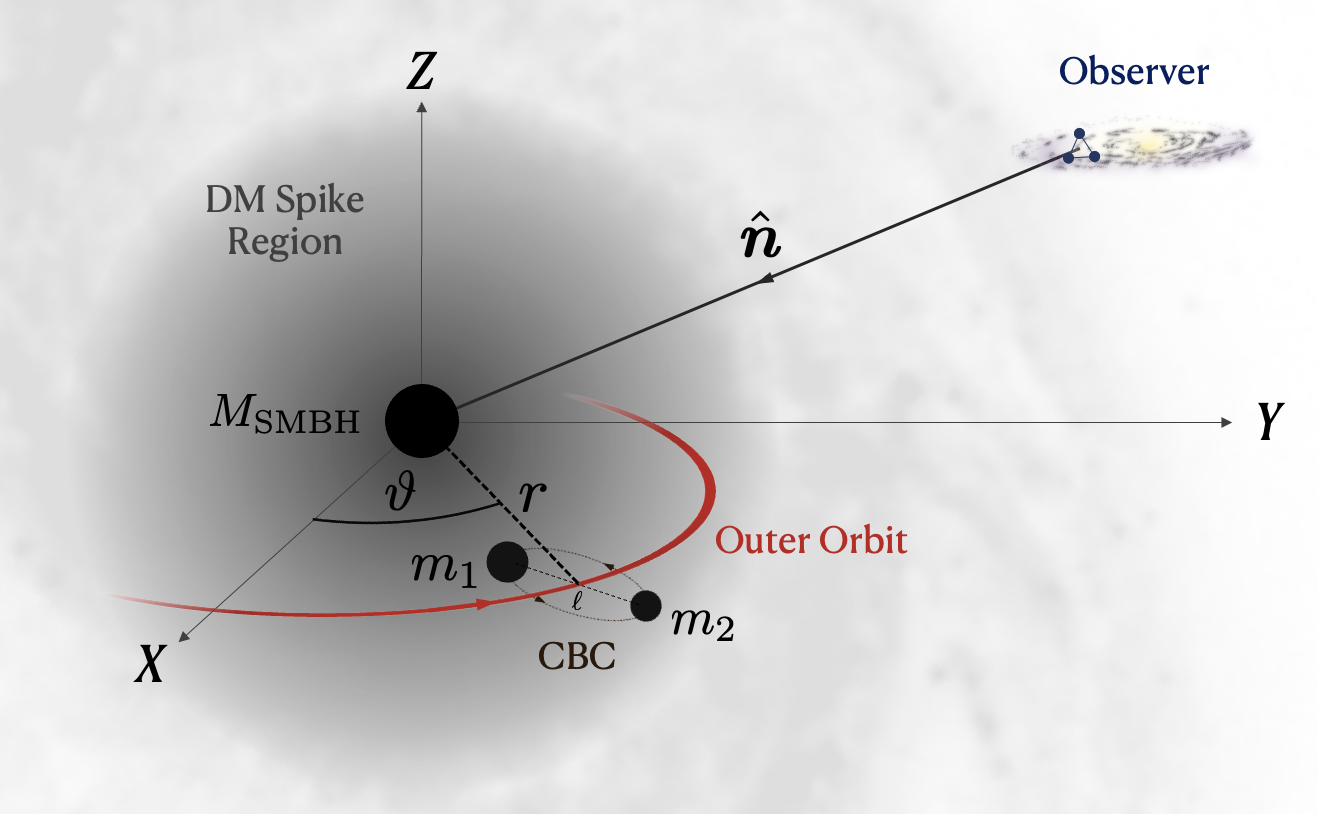}
  \caption{{\sc Schematic and Coordinates:} A compact binary system, with masses $m_1$ and $m_2$, orbiting around an SMBH of mass $M_{\rm SMBH}$, embedded in a dark matter spike shown as the spherical gray region inside the galactic halo.  The solid red curve marks the outer orbit of the CBC, i.e., the orbit of the CoM of the CBC, with angular momentum along the $Z$ axis and the angular coordinate of its CoM, $\vartheta$, measured counterclockwise from the $X$ axis. The observer lies in the $Y$–$Z$ plane and receives gravitational waves from the CBC from direction $\boldsymbol{n}$ at an angle $\iota_{\rm out} = {\rm arccos}(\hat{\boldsymbol{n}} \cdot \hat{z})$ from the $Z$ axis.
}
  \label{fig:coordsetup}
\end{figure}

\prlsec{Method.} 
We consider a CBC whose CoM executes a noncircular orbit through the DM halo around a supermassive black hole (SMBH) in a Milky Way-sized galaxy (See Fig.\,\ref{fig:coordsetup} for a schematic). The corresponding DM density is taken to have a spiky profile~\cite{Gondolo:1999ef},
\beq\label{CDM+BH}
\rho_{\rm DM}({r}< {r}_{\rm sp})= 
~\rho_{\rm sp}\left(\frac{r}{r_{\rm sp}}\right)^{-\gamma_{\rm sp}}\,,
\eeq
within the spike radius $r_{\rm sp}$,
beyond which it transitions to the initial unmodified profile $\rho_{\rm DM}(r > {r}_{\rm sp}) \propto r^{-1}$, where we adopt the Navarro-Frenk-White (NFW) form within the NFW scale-radius. Defining $\Sigma\equiv\rho_{\rm sp} r_{\rm sp}$, we treat $\Sigma$ and $\gamma_{\rm sp}$ as the parameters modeling the background density. The spike radius $r_{\rm sp}$ is determined via $r_{\rm sp}\approx0.2\,r_{2M}$, with $r_{2M}$ being the radius at which the unmodified NFW profile encloses a mass equal to $2M_{\rm SMBH}$, i.e., twice the SMBH mass. The inner radius of the spike may reach down to $4GM_{\rm SMBH}/c^2 \ll r_{\rm sp}$~\cite{Sadeghian:2013laa}.

The spike index $\gamma_{\rm sp}$ is sensitive to the underlying assumptions. 
Adiabatic accretion onto the SMBH predicts 
 $\gamma_{\rm sp} = (9-2\gamma_{\rm halo})/(4 - \gamma_{\rm halo})$, where $\gamma_{\rm halo}$ is the index of the initial DM halo profile~\cite{Gondolo:1999ef}; see ref.~\cite{Ullio:2001fb} for a derivation. An initial NFW profile with $\gamma_{\rm halo}=1$, chosen as our benchmark case, results in a steep spike with $\gamma_{\rm sp}=7/3$. 
However, $\gamma_{\rm sp}$ can be significantly smaller, e.g., if the galaxy is a result of a recent merger or due to the effect of star clusters, taking values between 1/2 and 7/3~\mbox{\cite{Milosavljevic:2001vi,Merritt:2002vj,Gnedin:2003rj,Kamermans:2024ieb}}. Additionally, baryonic contributions are expected to modify the density profile. We do not explicitly model this effect here -- instead, we explore a wide range of realistic $\gamma_{\rm sp}$ values to assess whether steepening beyond a canonical $\gamma_{\rm halo} = 1$ cusp is observationally testable.

Using the density profile in Eq.\,\eqref{CDM+BH}, the mass in the halo up to a radius $r<r_{\rm sp}$ is given by
\beq\label{eq: menc}
M_{\rm halo}(r)=
\frac{4\pi \rho_{\rm sp}r_{\rm sp}^{3}}{3-\gamma_{\rm sp}}\left(\frac{r}{r_{\rm sp}}\right)^{3-\gamma_{\rm sp}}.
\eeq
The orbit of a CBC through the spike region is governed by the gravitational potential $\Phi(r)$, which satisfies the Poisson equation, {$d\Phi/dr = G\left(M_{\rm SMBH}+M_{\rm halo}(r)\right)/r^{2}$}.

The GW signal from a source in relative motion is Doppler shifted. For constant velocity, this shift is fully degenerate with the source’s mass and cosmological redshift: a system with source-frame mass $M_{\rm s}$  and constant relative velocity is indistinguishable from the detector frame mass, $M_{\rm s} (1 + z_{\rm C}) (1 + z_{\rm D})$, for a static source, where $z_{\rm D}$ and $z_{\rm C}$ are the Doppler and cosmological redshifts, respectively.

On the other hand, a time-varying relative velocity, such as one due to a constant line of sight acceleration (LOSA) of the CoM, will produce modulations in the signal~\cite{Vijaykumar:2023waltzing,Bonvin:2016qxr,Yunes:2010zj,2024arXiv240715117T}. These modulations appear at $-4n$ Post-Newtonian (PN) orders in the phase of the emitted GWs when caused by $n^{\rm th}$ time-derivative of the line of sight velocity (LOSV) $v_{\rm L}$. For example, the modulations due to LOSA appear at $-4 \rm PN$, due to line of sight jerk (LOSJ) at $-8 \rm PN$, due to line of sight snap (LOSS) at $-16 \rm PN$, and so on. The derivatives of $v_{\rm L}$ furnish the relevant kinematic parameters
\begin{equation}
\Gamma_n \equiv \left. \frac{1}{n! \, c \, (1 + z_{\rm C})^n(1 + z_{\rm D})^n}\frac{d^n v_{\rm L}}{dt_{\rm o}^n}\right \vert_{t_{\rm o} = t_{\rm c}}\,,
\end{equation}
where $t_{\rm o}$ is the time in the observer's frame, and $t_{\rm c}$ is the time at coalescence. Then, under the assumption that $\vert \Gamma_n (t_{\rm o} - t_{\rm c})^n \vert \ll 1$, the GW waveform of the CBC moving with a time-varying relative velocity can be written as~\cite{2024arXiv240715117T}:
\begin{equation}
    \label{eq: WF}
    \tilde{h}_{\rm TV}(f) \propto \tilde{h}(f) \,e^{\, i{\scriptscriptstyle \sum\limits_{n=1}^{\infty}}\Delta \Psi_{-4n} (f;\Gamma_n)}\,,
\end{equation}
where $\tilde{h}(f)$ is the unmodulated GW waveform,
\begin{equation}
    \label{eq: pcnth}
    \Delta \Psi_{-4n} (f; \Gamma_n) = -\frac{ 2^{-8 n-7} 5^{n+1} u^{-8 n-5}}{(n+1) \nu^{n+1}} \left(-\frac{G M}{c^3}\right)^n \Gamma_n,
\end{equation}
$u \equiv (\pi G M f / c^3)^{1/3}$, $M \equiv m_1 + m_2$, $\nu \equiv m_1m_2/M^2$ is the symmetric mass ratio, $m_{1,2}$ are the detector-frame masses of the binary components, and $f$ is the observed GW frequency. The key physical effect is a time variation of the travel time of the GW signal, as the CoM accelerates, which is encoded in the phase.

The kinematic parameters $\Gamma_n$ can be related to the derivatives of the potential. We derive expressions for acceleration $\boldsymbol{a}$, jerk $\boldsymbol{j}$, and snap $\boldsymbol{s}$, and  project these kinematic parameters onto the line of sight $\hat{\bm{n}}$ (LOS), to obtain $\Gamma_1 = \boldsymbol{a} \cdot \hat{\bm{n}}/ [c (1 + z_{\rm C})(1 + z_{\rm D})]$, $\Gamma_2 = \boldsymbol{j} \cdot \hat{\bm{n}} / [2 c^2 (1 + z_{\rm C})^2(1 + z_{\rm D})^2]$, and $\Gamma_3 = \boldsymbol{s} \cdot \hat{\bm{n}} / [6 c^3 (1 + z_{\rm C})^3(1 + z_{\rm D})^3]$. See the Supplementary Material for the explicit expressions for $\boldsymbol{a}$, $\boldsymbol{j}$, and $\boldsymbol{s}$ derived from the potential $\Phi$.

\begin{figure*}
    \centering
    \includegraphics[width=0.6\linewidth]{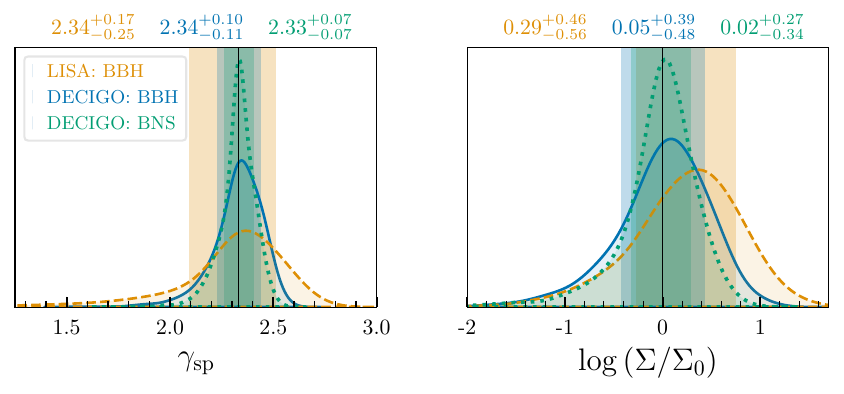}
    \caption{{\sc Profiling the DM Spike:} Expected 1d posteriors of the spike parameters $\gamma_{\rm sp}$ and $\Sigma$ for true values $\gamma_{\rm sp}=7/3\approx2.33$ and $\Sigma_0 = 1.07 \times 10^8 \, \rm M_{\odot} \, kpc^{-2}$. The dashed orange curves represent the posterior of the parameters for a future observation of an in-spike BBH by LISA, the solid blue and dotted green represent the same for a future observation of an in-spike BBH and BNS by DECIGO, respectively, while the shaded bands represent the corresponding 68\% CIs. See the text for the parameters of the binary systems and its environment.}
    \label{fig: 1d_post_LISA}
\end{figure*}

In the scenario considered, the GW signal is unaffected by dissipative effects. The acceleration due to DF is small, i.e., $a_{\rm DF}/a_{\rm sp}\lesssim 10^{-4}$, as is also the case in the IMRI scenario~\cite{Eda:2013gg,Eda:2014uha,Macedo:2013jja,Barausse:2014pra,Barausse:2014tra,Yue:2017iwc,Yue:2019ozq,Kavanagh:2020cfn,Coogan:2021uqv,Speeney:2022ryg,Tahelyani:2024cvk,Feng:2025fkc}. However, in contrast to the IMRI scenario, here the energy loss due to DF is also negligible compared to the GW loss, i.e.,\,\mbox{$\dot{E}_{\rm DF}/\dot{E}_{\rm GW}\lesssim 10^{-14}$}. This is mainly because the CBC is much more compact than the IMRI, and GW losses dominate. See the Supplementary Material for derivations of these estimates.

The environment and outer orbit are parameterized by $\{ \gamma_{\rm sp},\, \Sigma,\,M_{\rm SMBH},\,r, \,\cos \vartheta, \, v_r,\, v_t\}$, where $\vartheta$ is the angular coordinate of the binary in the outer orbital plane relative to the $X$-axis (see Fig.\,\ref{fig:coordsetup}), and $v_r$ and $v_t$ are the instantaneous radial and tangential speeds, respectively. We fix $\iota_{\rm out} = \pi/2$, i.e., assume an edge-on CBC. Note that the uncertainty in measurement of $\iota_{\rm out}$ is degenerate with mass-enclosed and $r$, but not $\gamma_{\rm sp}$. 
To obtain the constraints on the environment + outer orbit parameters, we perform a Fisher Matrix analysis on the kinematic parameters $\{\Gamma_1,\, \Gamma_2,\, \Gamma_3\}$, together with $\{ \ln D_{\rm L}, \, \ln \mathcal{M}, \, \ln \nu \}$, where $\mathcal{M} \equiv M \nu^{3/5}$ is the detector frame chirp mass of the binary, and then sample these parameters through the log-likelihood of the kinematic parameters using \texttt {emcee}~\cite{emcee} by setting uniform priors on all parameters. See Table\,\ref{tab: priors} in the Supplementary Material for details of the priors used in this study.

\prlsec{Results.} We consider three scenarios\,\footnote{Masses are mentioned in the source-frame.}:

\begin{itemize}[leftmargin=10pt]

\item[] {\bf{LISA-BBH}} \quad A  five-year observation of a $160\,\rm M_{\odot}$-$160\,\rm M_{\odot}$ BBH system at $100 \, \rm Mpc$ in LISA. Incorporating the cosmological redshift $z_{\rm C} \approx 0.033$ in mass-redshift degeneracy here as well sets the frequency range to be $5.75 \, {\rm mHz} - 1 \, \rm Hz$, which results in an optimal Signal-to-Noise Ratio\,(SNR) of $\approx 288$.

\item[] {\bf{DECIGO-BBH}} \quad A four-year observation of a $10\, \rm M_{\odot}$-$10\,\rm M_{\odot}$ BBH system at $1 \, \rm Gpc$ in DECIGO. Incorporating the cosmological redshift $z_{\rm C} \approx 0.198$ in mass-redshift degeneracy sets the frequency range to be $0.032 - 10\, \rm Hz$, which results in an optimal SNR of $\approx 1835$.

\item[] {\bf{DECIGO-BNS}} \quad A four-year observation of a $1.4\,\rm M_{\odot}$-$1.4\, \rm M_{\odot}$ binary neutron star (BNS) system at $1 \, \rm Gpc$ in DECIGO. Incorporating the cosmological redshift $z_{\rm C} \approx 0.198$ in mass-redshift degeneracy sets the frequency range to be $0.11 - 10\, \rm Hz$, which results in an optimal SNR of $\approx 334$.
\end{itemize}

\begin{figure*}
    \centering
\includegraphics[width=0.75\textwidth]{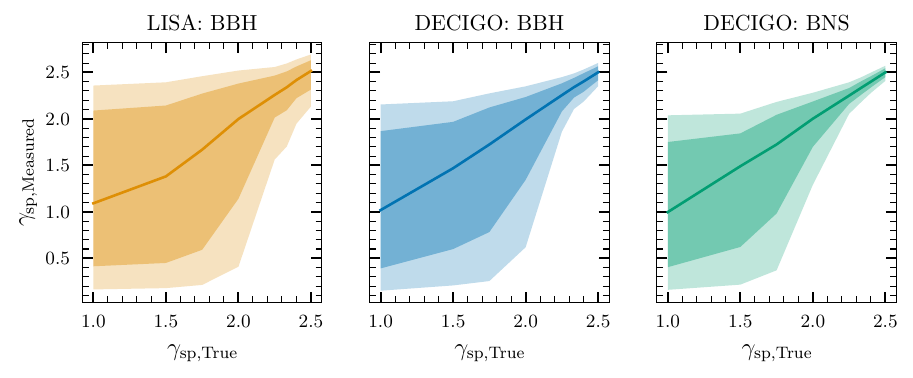}
    \caption{{\sc Precision in Spike Profiling: }Uncertainties in the measurement of $\gamma_{\rm sp}$ for a given value $\gamma_{\rm sp, True}$, for benchmark observations at LISA (left panel), and DECIGO (middle and right panels), as described in the text. The solid lines represent the median values, the dark bands represent the 68\% CIs, while the light bands represent the 90\% CIs.}
    \label{fig: gamma_prec_band}
\end{figure*}

We assume these binaries to be at a distance of $0.04 \, \rm pc$ from the center of a Milky Way-like galaxy, with a central BH of mass $M_{\rm SMBH} = 4 \times 10^6 \, \rm M_{\odot}$; note that the distance is significantly larger than the SMBH's Schwarzchild radius $\approx 38\,\mu{\rm pc}$ but well within the spike radius $\approx 20~{\rm pc}$. For these examples, we have set $z_{\rm D}=0$, i.e., there is no time-independent Doppler shift (which would be fully degenerate with source mass). We assumed $v_r=500 \, {\rm km/s}$, $v_t= 100 \, \rm km/s$ (see e.g., ref.\cite{2009A&A...502...91S}) and $\cos \vartheta = 0.5$. We consider a number of values $\gamma_{\rm sp}$ in the range $\left[1/2,\,7/3\right]$ to explore a wide range of scenarios proposed for DM spikes.

For these systems, the time variation is a small effect and is difficult to detect with the current generation of GW detectors.
The planned space-based detectors, such as LISA and DECIGO, will be able to monitor the CBCs for many years and be sensitive to subtle changes that could allow for a precise reconstruction of the density profile. 
For example, for DECIGO-BBH $\Gamma_1 \sim 10^{-12} \, {\rm s^{-1}}$, can be measured with precision $\Delta \Gamma_1 \sim 10^{-16} \, {\rm s^{-1}}$, which is about an order of magnitude larger than the change in LOSA due to the presence of a DM spike with $\gamma_{\rm sp} = 7/3$ around an SMBH. At the same time, $\Gamma_2 \sim 10^{-22}\,{\rm s^{-2}}$ can also be measured with precision $\Delta \Gamma_2 \sim 10^{-24} \, {\rm s^{-2}}$, which is comparable to the change in LOSJ due to the presence of a DM spike. This is confirmed upon examination of Fig.~\ref{fig: kp_corn} in the Supplementary Material. 

In Fig.\,\ref{fig: 1d_post_LISA}, we show the posteriors of $\gamma_{\rm sp}$ and $\Sigma$ inferred for the binary systems considered in LISA and DECIGO, for a model with true values $\gamma_{\rm sp} = 7/3$ and $\Sigma = \Sigma_0$, where $\Sigma_0 \equiv \rho_{\rm sp}^{\rm MW} r_{\rm sp}^{\rm MW} \simeq 1.07 \times 10^8 \, \rm M_{\odot} \, kpc^{-2}$. Notice that both parameters are well constrained within their 68\% credible intervals (CI), i.e., approximately $1\sigma$. 
For completeness, in the Supplementary Material, we also show the corner-plots for the kinematic parameters (Fig.\,\ref{fig: kp_corn}) and all environment + outer orbit parameters (Fig.\,\ref{fig: ful_env_p_lisa}). As seen from Fig.\,\ref{fig: 1d_post_LISA}, we can rule out not only the Kepler potential, i.e., $\Sigma = 0$, at $68\%$ level, but also other DM profiles such as the inner NFW with $\gamma_{\rm sp} = 1$ or even a prompt cusp with $\gamma_{\rm sp}=3/2$.

To explore the distinguishability of different DM spike profiles, we repeat the analysis for different values of $\gamma_{\rm sp}$, for the CBCs considered in LISA and DECIGO, by keeping all other parameters fixed. 
Fig.\,\ref{fig: gamma_prec_band} shows the $68\%$ and $90\%$ CI interval bands on the measurability of $\gamma_{\rm sp}$ as a function of the true value of $\gamma_{\rm sp,\,True}$. The measurability of $\gamma_{\rm sp}$ improves significantly for steeper DM spikes, reaching a few-percent--level precision for the largest values considered here. This is because a higher value of $\gamma_{\rm sp}$ will result in a larger enclosed mass and hence a deeper gravitational potential, which will lead to larger kinematic parameters. Therefore, the steeper the DM spike, the easier it will be to distinguish it from other DM spike profiles. 
It is also less degenerate with the Keplerian (and, potentially, baryonic) contributions. For completeness, the posteriors for different values of $\gamma_{\rm sp}$ are shown in the Supplementary Material (Fig.\,\ref{fig: Violin_plot_gamma}).

\prlsec{Discussion.} 
In this \emph{Letter}, we have presented a new method to profile DM spikes and measure the spike index. It relies on detecting the acceleration, jerk, etc., which are encoded as a time-varying modulation of the GW signal of a coalescing binary in orbit inside the spike, and is most effective for steep spikes with $\gamma_{\rm sp}\gtrsim 1.75$. 

The proposed approach paves a new way to study the nature of DM using its small-scale clustering, independent of detailed particle physics and insensitive to complex astrophysical processes. For the type of system considered, viz. stellar-mass binaries inside the spike, the GW modulation arises solely from the time-dependent CoM acceleration. Both DF-induced drag and energy loss are negligible. The compactness of the CBC yields GW losses far exceeding DF losses, unlike DF-driven IMRIs,
despite spike-induced accelerations mildly dominating in both cases (see Supplementary Material).
Unlike annihilation-based methods, which assume self-annihilating dark matter, our gravitational approach is largely insensitive to particle physics, except where it shapes the density profile. The proposed GW probe offers a clean and model-agnostic complement to the search for EM signatures. Moreover, this method is a novel gravitational probe of the innermost regions of galaxies. This is particularly true for a cosmological sample, where it is difficult to resolve the orbital motion of stars to study gravitational signatures. Given that the available evidence for dark matter is from its gravitational effects, such an approach likely presents a robust strategy. The method crucially relies on the existence of mergers within $\mathcal{O}\left(1\,{\rm pc}\right)$ of the center of galaxies. It has been argued that such mergers occur at a non-trivial rate (see, e.g., \cite{Fragione:2018yrb}), suggesting that several of them might be observed by future space based detectors.

In this work, we have not explicitly modeled the baryons. The gaseous component may be non-negligible; however, it is expected to have a flatter profile that is not degenerate with the spike. The stellar component can have a steep profile, like the collisionless DM, but as long as it is subleading, we expect our results to remain unaffected. Interplay between the DM and baryonic halos, e.g., stellar heating, can weaken the spike.
Here, we have consider the various values of $\gamma_{\rm sp}$ to account for such effects. In the future, we will explore if the method can distinguish between the two collisionless components.

\prlsec{Acknowledgments.}
We thank organizers and participants of the ``Pune-Mumbai Cosmology and Astroparticle Physics (PMCAP) Meeting'' held at the IIT Bombay in the Spring of 2025, during which this project was initiated. At TIFR, the work of PC and BD is supported by the Dept. of Atomic Energy (Govt. of India) research project RTI 4002. SJK acknowledges support from ANRF/SERB Grants SRG/2023/000419 and MTR/2023/000086. The work is partially supported by the Dept. of Science and Technology (Govt. of India) through a Swarnajayanti Fellowship to BD.

\prlsec{Software.} \texttt{NumPy} \citep{vanderWalt:2011bqk}, \texttt{SciPy} \citep{Virtanen:2019joe}, \texttt{astropy} \citep{2013A&A...558A..33A, 2018AJ....156..123A}, \texttt{Matplotlib} \citep{Hunter:2007}, \texttt{jupyter} \citep{jupyter}, \texttt{Numdifftools} \citep{numdifftools}.

\bibliography{References}

\clearpage

\onecolumngrid

\setcounter{equation}{0}
\setcounter{figure}{0}
\setcounter{table}{0}
\makeatletter
\renewcommand{\theequation}{S\arabic{equation}}
\renewcommand{\thefigure}{S\arabic{figure}}
\renewcommand{\thetable}{S\arabic{table}}

{\centering
\begin{center}{\large Supplementary Material\\[3ex] \textbf{Profiling Dark Matter Spikes with Gravitational Waves from Accelerated Binaries}}\end{center}
}
In this Supplementary Material, we provide derivations of some of the expressions used in the main text, the specifications assumed for LISA and DECIGO, adopted priors in the statistical analyses, and the complete posteriors of kinematic and environment\,+\,orbit parameters.

\section{Estimates of Dynamical Friction and Tidal Effects}

A compact body of mass $M$, moving through a local matter density $\rho$ experiences a force~\cite{chandrasekhar1943},
\begin{equation}
F_{\rm DF}\approx 4\pi G^2\rho\,M^2 N\ln\Lambda/v^2\,,
\end{equation}
where $N$ is the fraction of particles with speed less than $v$ and $\ln\Lambda$ is the Coulomb logarithm. In the case of CDM, $N\simeq 1$, and in this work we adopt $\ln\Lambda = 3$ \cite{Gualandris:2007nm}.
As the compact binary coalesces within the high-density spike around the SMBH, it loses energy via both GW emission and DF,
\beq
\dot{E} = \dot{E}_{\rm GW} + \dot{E}_{\rm DF}^{\rm CoM} + \dot{E}_{\rm DF}^{\rm Rot},
\eeq
where the DF terms represent contributions from drag on the binary’s center of mass (CoM) and its rotation (Rot) about the CoM through the medium.

A direct comparison between the GW and DF energy loss terms reveals that the former dominates in the regime of interest,
 \beq
  \frac{\dot{E}^{\rm CoM}_{\rm DF}}{\dot{E_{\rm GW}}}\approx 2.0\times 10^{-14} \left(\frac{\ell}{10^4\,{\rm km}}\right)^5\frac{(1+q^2)}{(1+q)q^2}\left(\frac{m_2}{30\,\rm M_{\odot}}\right)^{-3}\left(\frac{r/r_{\rm sp}}{10^{-3}}\right)^{-7/3}\left(\frac{250{\rm km/s}}{v_{\rm CoM}}\right),
 \eeq
and,
 \beq
 \frac{\dot{E}^{\rm Rot}_{\rm DF}}{\dot{E}_{\rm GW}}\approx 2.5\times 10^{-16}\left(\frac{\ell}{10^{4}~{\rm km}}\right)^{11/2}\frac{(1+q^{2})}{(1+q)^{3/2}q^2}\left(\frac{m_2}{30\,\rm M_{\odot}}\right)^{-7/2}\left(\frac{r/r_{\rm sp}}{10^{-3}}\right)^{-7/3},
 \eeq
 where $q = m_1/m_2$ is the mass ratio of the CBC and $\ell$ is the CBC separation. This is primarily because the GW energy loss scales steeply with the binary separation as $\ell^{-5}$, and $\ell$ is much smaller than the distance to the SMBH, i.e., $\ell \ll r$. Consequently, both $\dot{E}_{\rm DF}^{\rm CoM}$ and $\dot{E}_{\rm DF}^{\rm Rot}$ are subdominant to $\dot{E}_{\rm GW}$. Moreover, since the DF terms depend more weakly on radial separation, they remain approximately constant over the compact binary scale. Whereas previous literature has focused on IMRIs~\cite{Eda:2014uha,Macedo:2013jja,Barausse:2014pra,Barausse:2014tra,Yue:2017iwc,Yue:2019ozq,Kavanagh:2020cfn,Coogan:2021uqv,Speeney:2022ryg,Tahelyani:2024cvk,Feng:2025fkc}, where the binary separation is very large ($\ell\approx r$) and GW losses comparable or smaller than the DF losses, our approach --- dominated by GW emission --- leverages the kinematic signatures encoded in the GW signal of such CBCs as a clean probe of the inner structure of the spike.

The spike contribution to the acceleration of the CBC, $a_{\rm sp}$ is derived as $4\pi G\rho_{\rm sp}r^{1-\gamma_{\rm sp}}r_{\rm sp}^{\gamma_{\rm sp}}/(3-\gamma_{\rm sp})$~\cite{Eda:2013gg}. Hence, in the relevant zone, $r\ll r_{\rm sp}$, the dynamical friction is subdominant to the acceleration of the CBC due to the spike
\beq
\frac{a_{\rm DF}}{a_{\rm SP}} \approx 1.9\times 10^{-4}\left(\frac{v}{250~{\rm km/s}}\right)^{-2}(1+q)\left(\frac{m_2}{30~{\rm M_\odot}}\right)\left(\frac{r/r_{\rm sp}}{10^{-3}}\right)^{-1}\,,
\eeq
where, we show how this scales with respect to parameters appropriate for a Milky Way-like galaxy.

We may also consider the tidal forces exerted by the SMBH on the binary. Taking a conservative approach, we compute the maximum value of the tidal force, $F_{\rm TD}$, where the binary is oriented along the radial direction from the SMBH. A straightforward computation reveals that  the tidal force per unit mass $a_{\rm TD}$ is weaker, relative to the direct gravitational influence of the SMBH, by a factor proportional to 
\beq
\frac{a_{\rm TD}}{a_{\rm BH}}\sim 3\times 10^{-12}\left(\frac{\ell}{10^4~{\rm km}}\right)\left(\frac{r/r_{\rm sp}}{10^{-3}}\right)^{-1}.
\eeq%
In the parameter space of our interest, the energy lost by the GW emission is orders of magnitude larger than the energy lost due to tidal effects. These estimates demonstrate that dynamical friction or tidal pull have negligible effects on our measurement, as claimed in the main text.

\section{Expressions for the Kinematic Parameters}\label{sec: kip_par_der}

The radial acceleration of a point mass, due to the black hole and halo potential, can be calculated as \cite{2024arXiv240715117T}
\beq
\boldsymbol{a} &= -\frac{d\Phi}{d{\boldsymbol{r}}},
\eeq
where $\Phi$ is evaluated by solving the relevant Poisson equation, $d\Phi/dr = G\left(M_{\rm SMBH}+M_{\rm halo}(r)\right)/r^{2}$, with the $M_{\rm halo}(r)$ given in Eq.~\eqref{eq: menc} in the main text. Therefore, within the spike profile
\beq\label{eq: acc}
a_r  =- \frac{GM_{\rm SMBH}}{r^2} - 4\pi G \rho_{\rm sp}\frac{r^{1-\gamma_{\rm sp}}}{3-\gamma_{\rm sp}}r_{\rm sp}^{\gamma_{\rm sp}},
\eeq
where, for a purely radial force, the angular component $a_\vartheta$ vanishes. The jerk $\boldsymbol{j}$ is evaluated as
\beq
~&\boldsymbol{j} = \left(\frac{da_r}{dt} - a_\vartheta\Omega\right)\mathbf{e}_r+\left(\frac{da_\vartheta}{dt}+a_r\Omega\right)\,\mathbf{e}_\vartheta,
\eeq
where $\Omega\equiv v_t/r$ is the angular speed. The radial ($j_r$) and the tangential components ($j_t$) are given as the following
\beq\label{eq: jerk}
~j_r  &= \frac{da_r}{dr}v_{r} =  \left[\frac{2GM_{\rm SMBH}}{r^3} - 4\pi G \rho_{\rm sp}\frac{1 - \gamma_{\rm sp}}{3-\gamma_{\rm sp}}\left(\frac{r_{\rm sp}}{r}\right)^{\gamma_{\rm sp}}\right]v_{r},\\
~j_\vartheta& = v_t a_r/r.
\eeq
Similarly, the snap $\boldsymbol{s}$ is calculated as
\beq
~&\boldsymbol{s} = \left(\frac{dj_r}{dt} - j_\vartheta\Omega\right)\mathbf{e}_r+\left(\frac{dj_\vartheta}{dt}+j_r\Omega\right)\,\mathbf{e}_\vartheta,
\eeq
and the components of snap are found to be
\beq\label{eq: snap}
s_r &= -\frac{2 G^2 M_{\rm \rm SMBH}^2}{r^5} - \frac{3 G M_{\rm SMBH} }{r^4}(2v_r^2-v_t^2) + \frac{4\pi G^2 (1+\gamma_{\rm sp}) M_{\rm SMBH}\rho_{\rm sp}r_{\rm sp}^{\gamma_{\rm sp}}}{\gamma_{\rm sp} - 3}r^{-\gamma_{\rm sp}-2}\\
&~~~~~~~ + \frac{4\pi G\gamma_{\rm sp}\rho_{\rm sp}r_{\rm sp}^{\gamma_{\rm sp}}}{\gamma_{\rm sp} - 3}r^{-1-\gamma_{\rm sp}}\left((\gamma_{\rm sp}-1)v_r^2-v_t^2\right) + \frac{16\pi^2 G^2 (1 - \gamma_{\rm sp}) \rho_{\rm sp}^2r_{\rm sp}^{2\gamma_{\rm sp}}}{\left(\gamma_{\rm sp}-3\right)^2}r^{1-2\gamma_{\rm sp}} \,,\\
s_{\vartheta} &= \frac{2 G v_r v_{t}}{r}  \left(\frac{3 M_{\rm SMBH}}{r^3}-\frac{4 \pi  \gamma_{\rm sp}  \rho_{\rm sp}r_{\rm sp}^{\gamma_{\rm sp}} r^{-\gamma_{\rm sp} }}{\gamma_{\rm sp} -3}\right) \,.
\eeq

We then compute the kinematic parameters,
\begin{equation}
\Gamma_1 = \frac{\boldsymbol{a} \cdot \hat{\bm{n}}}{c(1 + z_{\rm C})(1 + z_{\rm D})}, \quad
\Gamma_2 = \frac{\boldsymbol{j} \cdot \hat{\bm{n}}}{2c^2(1 + z_{\rm C})^2(1 + z_{\rm D})^2}, \quad
\Gamma_3 = \frac{\boldsymbol{s} \cdot \hat{\bm{n}}}{6c^3(1 + z_{\rm C})^3(1 + z_{\rm D})^3}\,,
\end{equation}
that encode the projection of $\boldsymbol{a}, \boldsymbol{j},\boldsymbol{s}$, on the LOS direction, $\hat{n}$. The factors of $(1+z_{\rm C})(1+z_{\rm D})$ account for the combined effects of cosmological expansion and the LOS Doppler shift.

\section{Detector Sensitivity and S/N}

Assuming the unmodulated GW waveform $\tilde{h}(f)$ to be the form $\tilde{h}(f) = \mathcal{A} f^{-7/6} e^{i \Psi (f)}$~\cite{Buonanno:2009zt}, for a GW signal $\tilde{h}_{\rm TV}(f)$ of the form Eq.\,\ref{eq: WF}, the optimal SNR \cite{Robson_2019} can be written as:
\begin{equation}
    \varrho = \sqrt{4  \int_{f_l}^{f_h} \frac{\vert \tilde{h}_{\rm TV}(f) \vert ^ 2}{S_n (f)}df} = \sqrt{4 |\mathcal{A}|^2 \mathcal \int_{f_l}^{f_h} \frac{f^{-7/3}}{S_n(f)} df} \mathcal{Q}(\iota)
\end{equation}
where $\mathcal{A} \equiv \pi^{-2/3}(5/24)^{1/2}(c/ D_{\rm L})\left(G\mathcal{M}/c^3\right)^{5/6} $, $\mathcal{M}$ is the detector frame chirp mass, $D_{\rm L}$ is the luminosity distance, $f_l$ and $f_h$ are the lower and upper cut-off frequencies, $S_n(f)$ is the one-sided noise power spectral density (PSD), and
\begin{equation}
    \mathcal{Q}(\iota) = \sqrt{\left( \frac{1 + \cos^2\iota}{2}\right)^2 + \cos^2\iota}\,,   
\end{equation}
where $\iota$ is the inclination of the (inner) binary relative to the LOS. We assume face-on inner orbits, which maximizes the SNR. When the binary is inclined, the SNRs will be reduced by a factor $\mathcal{Q} (\iota)/\mathcal{Q} (\iota = 0)$.

To estimate the precision with which the source parameters can be extracted, we employ a Fisher matrix analysis \cite{CutlerFlanagan}. Here, the GW likelihood is assumed to peak at the true values of the source parameters, and its shape in the vicinity of the peak is approximated to be a multivariate Gaussian defined by a covariance matrix. The covariance matrix -- which contains the r.m.s errors on the environment and outer orbit parameters that measure the precision of their extraction -- is simply the inverse of the Fisher matrix. We refer the reader to \cite{2024arXiv240715117T}, in particular the Supplementary Material therein, for a detailed exposition of the method.

The noise PSDs ($S_n(f)$) for LISA and DECIGO are taken \cite{Robson_2019} and from \cite{PhysRevD.83.044011, PhysRevD.95.109901}. We assume $5$ and $4$ years of observation time and sensitivity bands $ [10^{-4}, 1]~{\rm Hz}$ and $[10^{-2}, 10]~{\rm Hz}$, respectively. The lower frequency $f_l$ used for evaluating the optimal SNR is set to ensure that the duration of the signal from that frequency to the upper limit of the frequency band is exactly equal to the observation time. 
The SNRs reported in the main text can be readily computed using the specifications provided above.

\section{Benchmark Values and Priors}\label{sec: priors}
We use uniform priors on all environment + outer orbit parameters. Table \ref{tab: priors} explicitly shows the benchmark values of all parameters used, and the priors assumed for sampling of environment + outer orbit parameters.
\begin{table}[!h]
    \centering
    \begin{tabular}{l l l}
        \toprule
        \textbf{Parameter\quad} & \textbf{Benchmark Value(s)\quad} & \textbf{Prior} \\
        \midrule
        $\gamma_{\rm sp}$ & $\{1,\,1.5,\,1.75,\,2,\,2.25,\,2.33,\,2.4,\,2.5\}$\mbox{\phantom{m}} & $\mathcal{U} (0.05,\, 2.95)$ \\
        $\Sigma$ & $\Sigma_0 \equiv 1.07 \times 10^8 \, {\rm M_\odot} /{\rm kpc}^2$& $\mathcal{U} (10^{-5},\, 50)$ \\
        $M_{\rm SMBH}$ & $4\times 10^6\,\rm M_\odot$ & $\mathcal{U} (0.1,\, 10)$ \\
        $r$ & 0.04\,{\rm pc}& $\mathcal{U} (0.004,\, 0.4)$ \\
        $\cos \vartheta$ & 0.5 & $\mathcal{U} (0,\, 1)$ \\
        $v_r$ & 500 km/s& $\mathcal{U} (0,\, 10)$ \\
        $v_t$ & 100 km/s& $\mathcal{U} (0,\, 10)$ \\
        \bottomrule
    \end{tabular}
    \caption{{\sc Priors Used During the Sampling of the Environment + Outer Orbit Parameters}: The priors $M_{\rm SMBH}$, $r$, and $\Sigma$ are in units of $10^6 \, \rm M_{\odot}$,  $\rm pc$, and $\Sigma_{0}$, respectively, while the priors on $v_r$ and $v_t$ are in units of $100\, \rm km/s$. $\mathcal{U}$ refers to a uniform distribution over the stated interval.}
    \label{tab: priors}
\end{table}

\section{1d and 2d Posterior Distributions}
\label{sec: example_corn_pl}
\begin{figure*}[htbp]
    \centering
\includegraphics[width=0.975\linewidth]{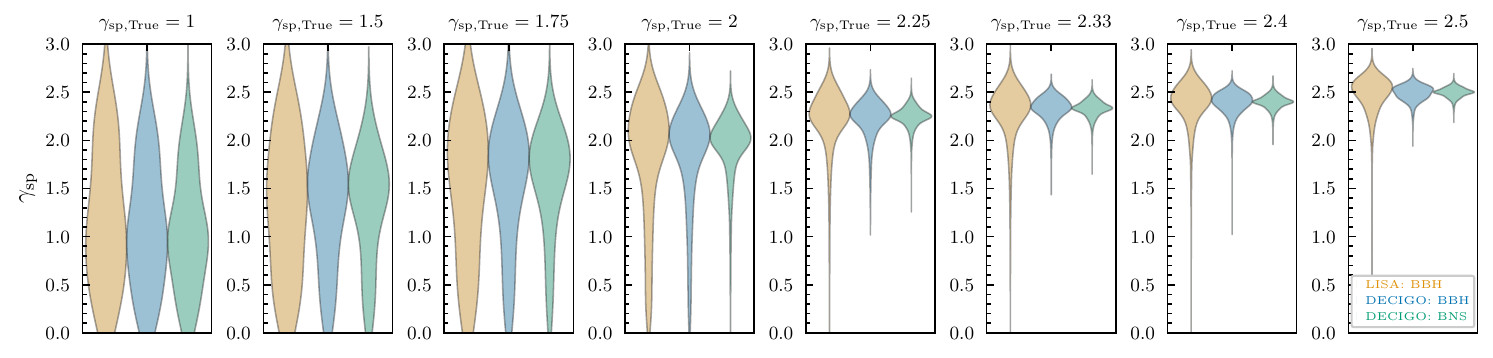}
    \caption{{\sc Violin Plots of 1d Posteriors for Different Spike Index:}  The orange, blue, and green violins represent the $1$d posteriors of $\gamma_{\rm sp}$ for {LISA-BBH}, {DECIGO-BBH}, and {DECIGO-BNS} systems, respectively, for the true spike index denoted by the top-label in each panel. The precision reaches to a few-$\%$ level for larger values of true $\gamma_{\rm sp}$.}
    \label{fig: Violin_plot_gamma}
\end{figure*}

In Fig.\,\ref{fig: Violin_plot_gamma}, we show the $1$d posteriors of $\gamma_{\rm sp}$ for different values of $\gamma_{\rm sp} \in \{1, \, 1.5, \, 1.75, \, 2, \, 2.25,\, 2.33,\, 2.4,\,2.5\}$, for the LISA and DECIGO systems in the main text, by keeping all other parameters unaltered.

\begin{figure}[htbp]
\centering
\includegraphics[width=0.45\linewidth]{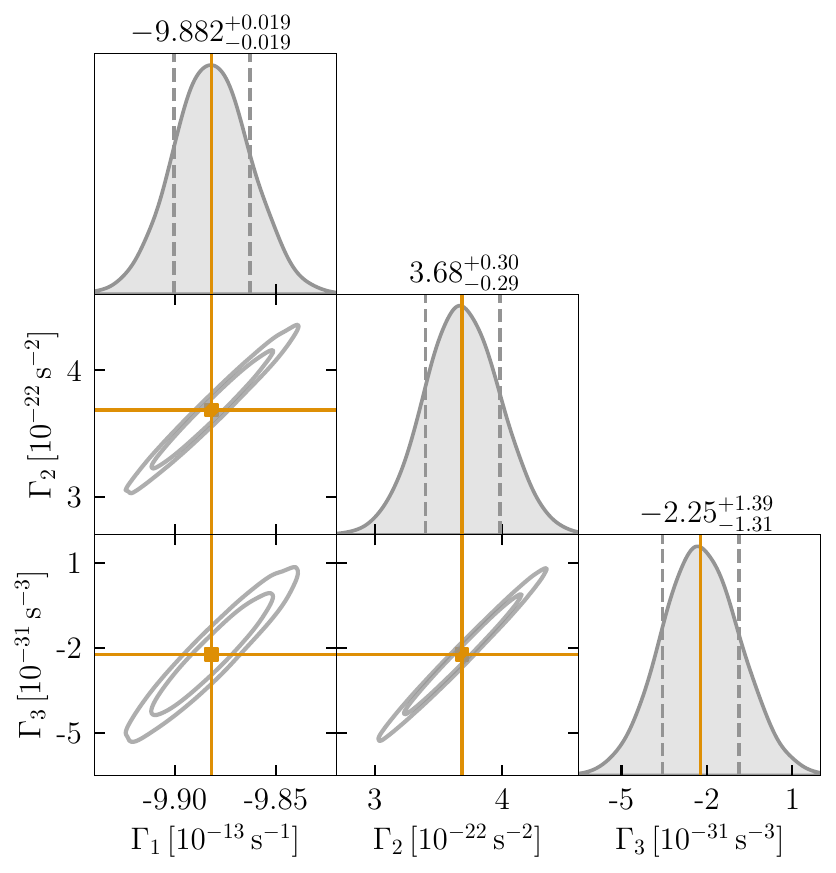}\\[5ex]
\includegraphics[width=0.45\linewidth]{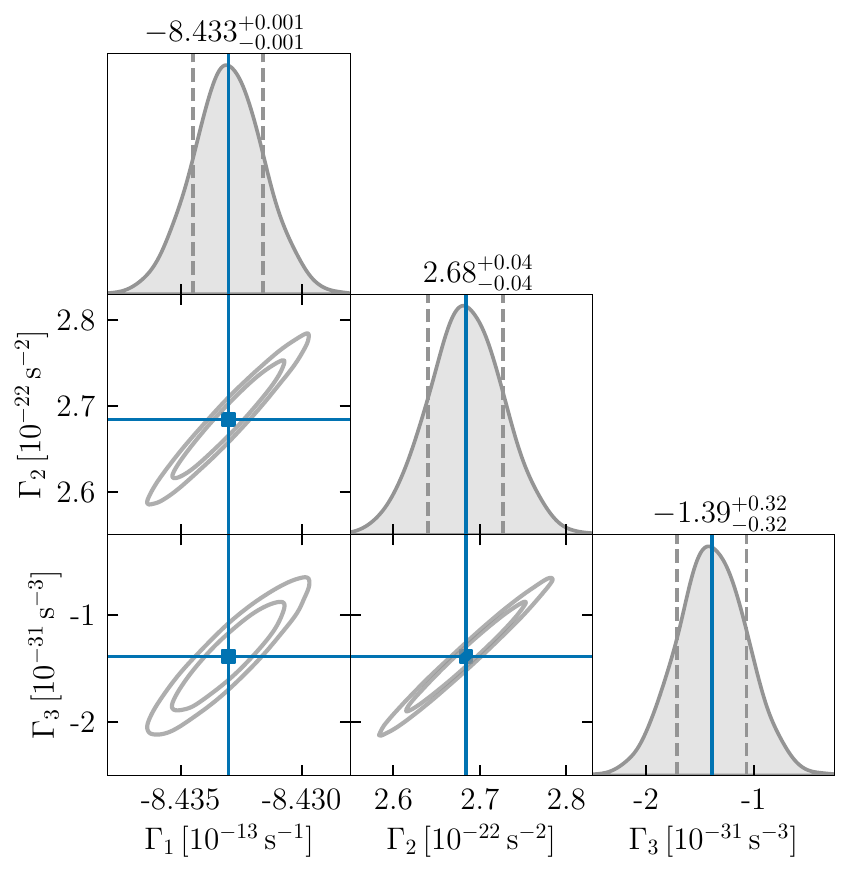}\qquad\includegraphics[width=0.45\linewidth]{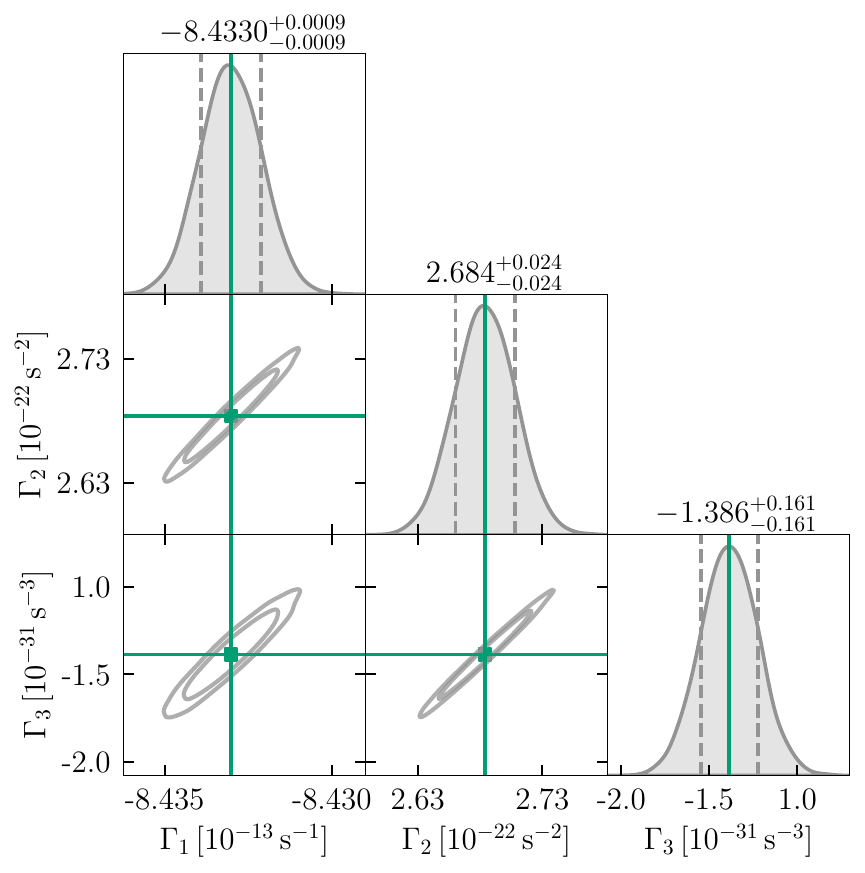}
    \caption{{\sc Corner Plots of Kinematic Parameters:} For the {LISA-BBH} (top), {DECIGO-BBH} (bottom-left), and {DECIGO-BNS} (bottom-right) systems considered in the main text, posteriors of LOSA ($\Gamma_1$), LOSJ ($\Gamma_2$), and LOSS ($\Gamma_3$) are shown. Orange/blue/green lines represent the true values of the parameters, gray dashed lines represent the 68\% CIs, gray contours represent the 68\% and 90\% CIs, while titles specify the mean and 68\% CI values.}
    \label{fig: kp_corn}
\end{figure}

\begin{figure*}
\centering
\includegraphics[width=0.95\linewidth]{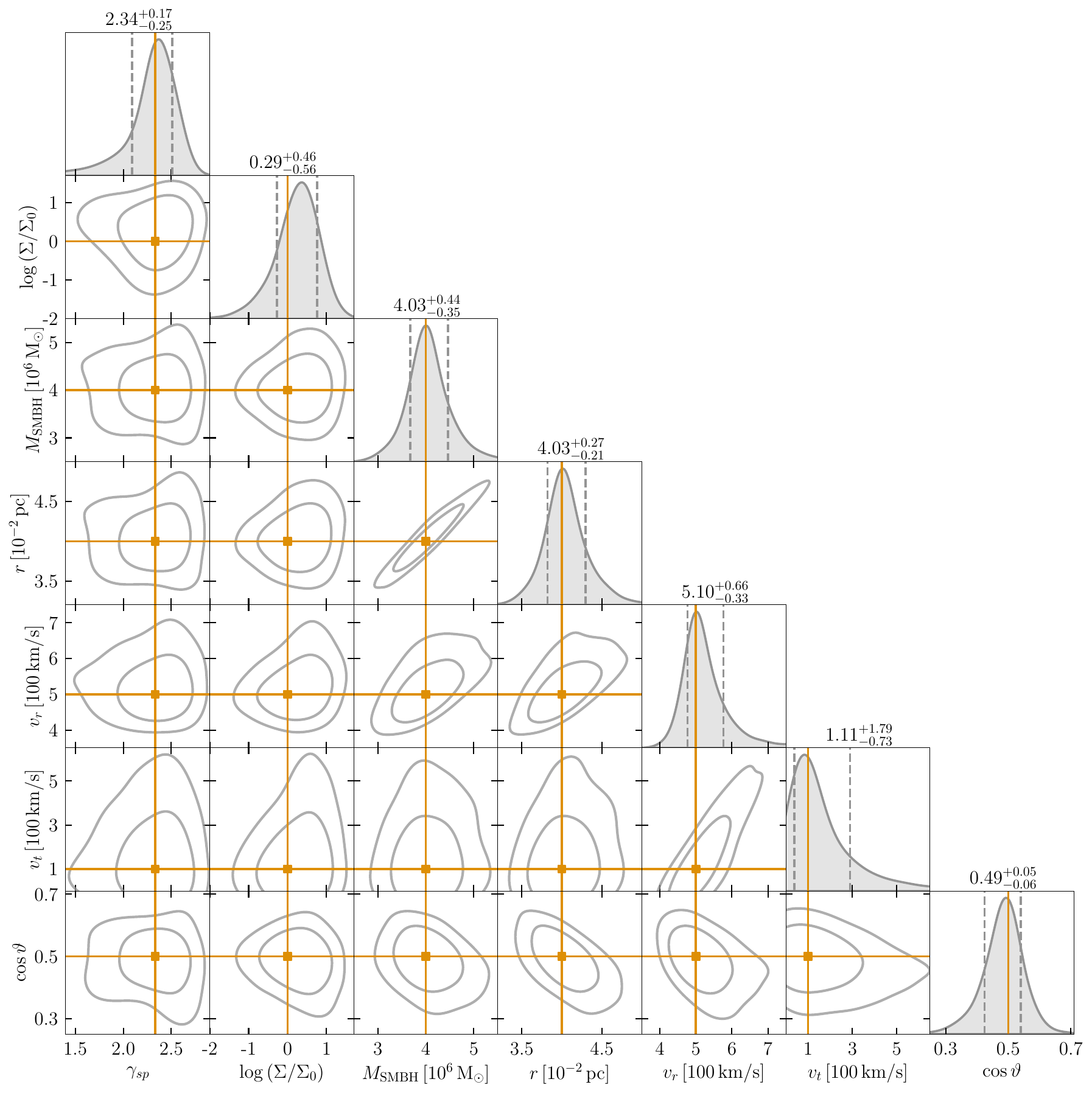}
    \caption{{\sc Corner Plot of the Environment + Outer Orbit Parameters for {LISA-BBH}}: Orange lines represent the true values of the parameters, gray dashed lines represent the 68\% CIs, gray contours represent the 68\% and 90\% CIs, while titles specify the mean and 68\% CI values.}
    \label{fig: ful_env_p_lisa}
\end{figure*}

In Figs.\,\ref{fig: kp_corn} and \ref{fig: ful_env_p_lisa}, we show the 1d and 2d posteriors of the kinematic and all environment + outer orbit parameters for the LISA and DECIGO systems, for which the $\gamma_{\rm sp}$ and $\Sigma$ were shown in Fig.\,\ref{fig: 1d_post_LISA} in the main text.

\begin{figure*}
\centering
\includegraphics[width=0.95\linewidth]{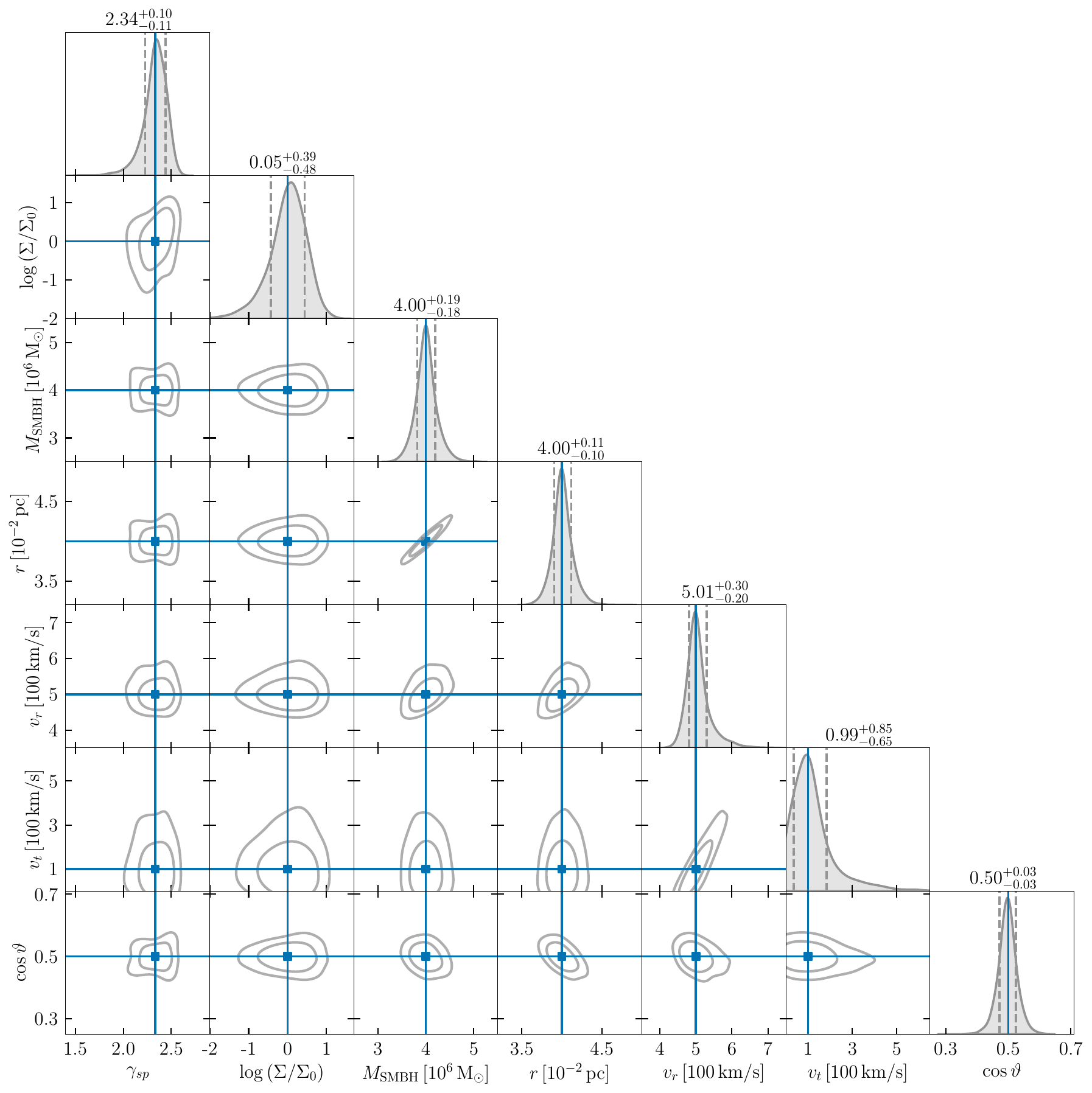}
    \caption{{\sc Corner Plot of the Environment + Outer Orbit Parameters for {DECIGO-BBH}}: Blue lines represent the true values of the parameters, gray dashed lines represent the 68\% CIs, gray contours represent the 68\% and 90\% CIs, while titles specify the mean and 68\% CI values.}
    \label{fig: ful_env_p_decigo}
\end{figure*}

\begin{figure*}
\centering
\includegraphics[width=0.95\linewidth]{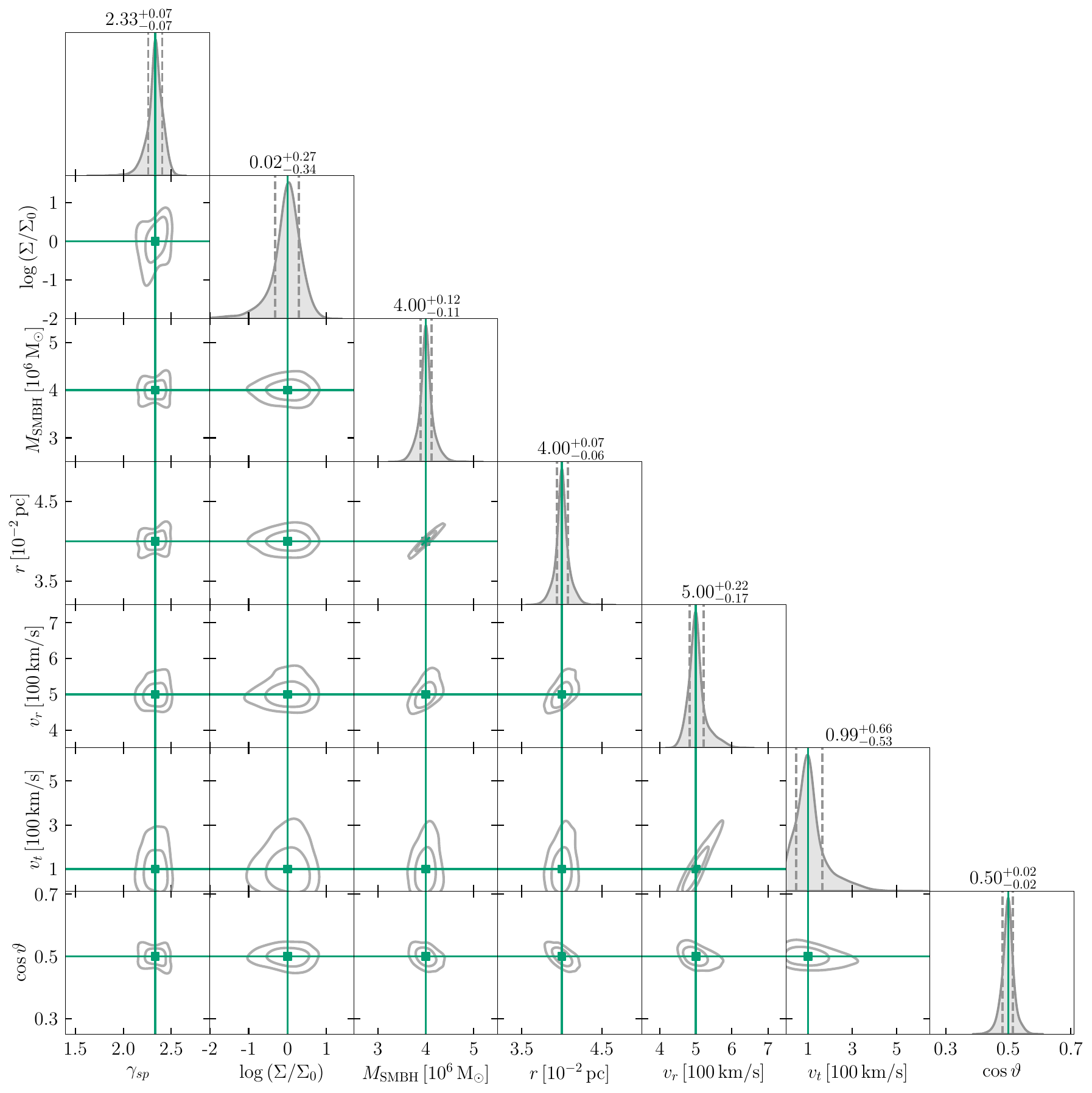}
    \caption{{\sc Corner Plot of the Environment + Outer Orbit Parameters for {DECIGO-BNH}}: Green lines represent the true values of the parameters, gray dashed lines represent the 68\% CIs, gray contours represent the 68\% and 90\% CIs, while titles specify the mean and 68\% CI values.}
    \label{fig: ful_env_p_decigo_bns}
\end{figure*}

\end{document}